\documentstyle[psfig,aps]{revtex}
\begin{document}
\draft
\preprint{MKPH-T-01-01}
\title{
\hfill{\small {\bf MKPH-T-01-01}}\\
{\bf
The \mbox{\boldmath$\eta NN$}-system at low energy within a three-body
approach\footnote{Supported by the Deutsche Forschungsgemeinschaft (SFB 443)}}
}
\author{A.\ Fix\footnote
{Permanent address: Tomsk Polytechnic University, 634034 Tomsk, Russia}
and H.\ Arenh\"ovel}
\address{
Institut f\"{u}r Kernphysik, Johannes Gutenberg-Universit\"{a}t,
       D-55099 Mainz, Germany}
\date{\today}
\maketitle

\begin{abstract}
The role of the $\eta NN$-interaction is studied in the low energy regime in
$\eta$-deuteron reactions as well as in coherent and incoherent
$\eta$-photoproduction on the deuteron using a three-body model with
separable two-body interactions.
The three-body approach turns out
to be quite essential in the most important lowest partial wave.
Results are presented for differential and total cross sections as well as
for the $\eta$-meson spectrum. They differ significantly
from those predicted by a simple rescattering model in which only first-order
$\eta N$- and $NN$-interactions in the final state are considered.
The major features of the experimental data of $\eta$-photoproduction
in the near-threshold region are well reproduced.
\end{abstract}

\section{Introduction}
The basic phenomenon, which plays a crucial dynamical role in
medium energy physics, is the excitation of baryon resonances
inside a nucleus, either in a hadronic or an electromagnetic reaction.
The phenomenological framework, in which the nuclear
dynamics in this energy region is described, is in terms of a series
of excitations and subsequent decays of baryon
resonances in a nuclear medium with intermediate meson exchange.

A particularly interesting topic is the field of $\eta$-meson physics,
because the $\eta N$-interaction is dominated at low energy by the almost
exclusive coupling to the $N^*$(1535)-resonance. Thus the $\eta$-nucleus
dynamics is shifted to that of the $N^*$(1535), i.e., to the $N^*$-nucleus
interaction. The essential question is, how well do we understand this
interaction. Until now, the $N^*$-nucleus dynamics has been treated
either purely phenomenological within the $N^*$-hole model~\cite{Oset,Kohn}
or by combining microscopic and phenomenological tools in the so called
local density approximation~\cite{Carr,Kripp} or using the BUU
model~\cite{Hombach}.  However, such approaches leave open the question of
the underlying ``elementary'' $N^*N$-interaction in few body systems, where
one must go beyond the simplified approaches for the $N^*$-dynamics in
nuclear matter. The present investigation is an attempt to shed some light
on this particular problem by studying elastic and inelastic
$\eta$-deuteron scattering as well as $\eta$-photoproduction on the
deuteron.

Elastic $\eta d$-scattering at low energy has been considered recently
in~\cite{Ued91,Wycech,Shev,Pena}, mainly in connection with a search for
rather exotic bound $\eta d$-states.
The outstanding feature of the $\eta d$-interaction, found in all analyses,
is its strong energy dependence which manifests itself
in the sharp enhancement of the cross section near zero energy.
The origin of this feature is a pole in
the $s$-wave scattering amplitude near the physical region. However,
despite all theoretical efforts, using rather nontrivial models, even a
qualitative understanding of the low-energy $\eta d$-interaction has not
been reached.  In fact, the different results show a strong qualitative
disagreement which cannot be explained alone by the large uncertainties in
our knowledge of the $\eta N$ low-energy interaction.  For example,
in~\cite{Shev,Shev2} the existence of an $\eta d$ s-wave resonance was
claimed, which, however, was not observed by other authors
\cite{Pena,Wycech3,FiAr00}. The authors of~\cite{Pena} have
found that weakening the strength of the $\eta N$-forces implies
the disappearance of the quasibound state without generating an $\eta
d$-resonance, while the strong enhancement of the cross
section remains.
Recently, we have shown in~\cite{FiAr00} that it is very
likely, that the strong energy dependence of the $\eta d$-scattering
amplitude is generated by a virtual state into which the bound state turns
by weakening the $\eta N$-interaction strength.  We also would like to
point out that our conclusion agrees with the one given in~\cite{Wycech3},
where the three-body equations have been solved by the partial summation of
the multiple scattering series.

Also the role of pion exchange in
the $N^*N$-interaction remains unclear. The results shown in
\cite{Pena2} support a very strong damping of the attraction in the $\eta
d$-system due to $\pi$-exchange. On the other hand, in
\cite{FiAr00} this effect was found to be insignificant. Therefore, the
physics of low-energy $\eta d$-interaction is by no means understood,
and further careful studies are required.

In the present paper we will continue and extend our study of the
$\eta NN$-system of~\cite{FiAr00,FiArPL}.
It is reasonable to expect that the virtual poles in the $\eta
NN$-scattering matrix found in \cite{FiAr00}
would have important consequences for elastic and
inelastic $\eta d$-scattering in the energy region of a few MeV above
threshold. This is the region which we wish to explore in this
paper, and it is our purpose to survey those important
features of the $\eta NN$-interaction, which are expected to influence the
$\eta$-production reactions. We would like to note that due to
rather large uncertainties of the experimental input on the one hand and a very
strong model dependence on the theory side, a considerable freedom remains in
fitting the isobar ansatz for the $\eta N$-$\pi N$-$\pi\pi N$ coupled channel
system to the low energy data. As a consequence, the results for the $\eta
N$-scattering length range from $a_{\eta N}=(0.27+0.22\,i)$~fm~\cite{Bh} to
$a_{\eta N}=(0.88+0.27\,i)$~fm~\cite{Bat}.
In this connection we think that it is very hard at present
to make precise quantitative statements about the character of the
$\eta d$ low-energy interaction based on the coupled channel approach alone.
Therefore, we will focus our attention primarily on the
qualitative aspects of this phenomenon which, we believe, are more or less
independent of future refinements of the $\eta N$-dynamics, when more
theoretical and experimental material will be available.

Information obtained from  photon-induced $\eta$-production
may be considered as complementary to that from $\eta d$-scattering.
Indeed, due to the large momentum transfer,
the $\eta$-photoproduction cross section is closely connected
with the high momentum part of the $N^*N$-interaction, which, as we will see,
is only of minor importance in the low-energy $\eta d$-scattering.
Photoproduction reactions
$\gamma d\to\eta d$ and $\gamma d\to\eta np$ have been extensively
studied over recent years in~\cite{Shev2,Krusche2,Metag,Ued92,Kamal97,Hoff}.
The interest in these reactions is motivated by two main
reasons. Firstly, one has the possibility to extract information about the
$\eta$-photoproduction strength on a neutron within the framework of a
detailed microscopic approach. This
question is closely connected with the problem of the isotopic
separation of the $N^*$-photoexcitation amplitude
discussed, e.g., in~\cite{Hoff,Breit,FiAr97,RiAr00}.
Secondly, these reactions permit a nondirect investigation of the
$N^*N$-interaction in different spin-isospin channels.
In the spirit of our theoretical focus, we will pay our attention mostly
to the second aspect.

Thus we will consider in this work the $N^*N$-interaction as an
$\eta NN$ three-body problem.
We would like to note, that the study of
the $N^*N$-interaction in analogy to the familiar $NN$ two-body potential
scattering, i.e., treating the $N^*$ essentially as a
stable particle, would be incorrect. This is because of large
retardation effects expected, e.g., in the $\eta NN$ channel, to which the
$N^*N$-system is coupled, leading to a strong nonlocality of the
$N^*N$-interaction. Therefore, any realistic approach
must include explicitly meson degrees of freedom. This aspect may
unambiguously be taken into account in the framework of a
three-particle model, where special theoretical techniques are
available for solving the appropriate dynamical equations.

Before dealing with the main subject, we describe in Sect.~\ref{formalism}
the essential ingredients of the formalism used in the present study.
The three-body equations are simplified without loss of essential physics
by using separable potentials for the two-body forces. We also pay some
attention to the explicit inclusion of the absorptive $\pi NN$ channel.
We then specialize in Sect.~\ref{scattering} to the problem of
$\eta d$-scattering which, although experimentally not accessible, supplies the
important input for the subsequent treatment of $\eta$-production.
In Sect.~\ref{unitarity} the consequences on the
break-up reactions arising from unitarity are considered.
Photoproduction is treated in Sect.~\ref{photoproduction} where we will
compare the present calculation to recent experimental data for the
coherent and incoherent reactions
$\gamma d\to\eta d$ and $\gamma d\to\eta np$.
We will show, that the main features of the $\eta
NN$-system, predicted by our model are in rather good agreement with
these measurements. The last Section~\ref{conclusion} is devoted to
conclusions. Here, we will also address some questions concerning the
$N^*$-nuclear dynamics and possible mechanisms of $\eta$-meson production
on heavier nuclei in the near-threshold region.

\section{Formalism}\label{formalism}

In this section we briefly review the basic features of three-body
techniques for the  application to the $\eta NN$-dynamics. The system
under consideration consists of two nucleons, $N_1$ and $N_2$, and an
$\eta$-meson, which will be denoted as particle 1,2 and 3, respectively.
In the c.m.\ frame the basic free particle states
$|\vec{p}_i,\vec{q}_i \rangle$ will
be characterized as usual by a pair of vectors $\vec{p}_i$ and
$\vec{q}_i$, where $\vec{p}_i$ is the relative momentum of a $(jk)$-pair
($j\neq i\neq k$) and $\vec{q}_i$ denotes the relative momentum of the unpaired
particle $i$ with respect to the c.m.\ frame of the pair.

In order to approximate the three-body equations in such a way that they become
practically solvable, it is customary to introduce a separable
ansatz for each two-body interaction. In our case this approximation has
also a physical motivation because of the strong dominance of the $s$-wave
pole terms in the low-energy $\eta N$- and $NN$-scattering matrices.  Thus
we will assume that the two-body driving forces can be approximated by
rank-one potentials, which, when regarded as operators
in the three-body Hilbert space, have the form
\begin{equation}
v_i=\gamma_i\int\frac{d^3q}{(2\pi)^3}
|i,\vec{q}\,\rangle \langle i,\vec{q}\,|\,,
\end{equation}
with $i$ being the channel index, in detail
\begin{equation}
|i,\vec{q}\,\rangle=\left\{
\begin{array}{lcl}
|N_1(\vec{q}\,),(N_2\eta)\rangle & \mbox{for} & i=1 \\
|N_2(\vec{q}\,),(N_1\eta)\rangle & \mbox{for} & i=2 \\
|\eta(\vec{q}\,),(N_1 N_2)\rangle & \mbox{for} & i=3
\end{array}\right\}\,.\label{channel}
\end{equation}

Here the ket $|i,\vec{q}\,\rangle =|i\rangle\otimes |\vec{q}\,\rangle$ is
defined such that
\begin{equation}
\langle \vec{p},\vec{q} \,| i,\vec{q}^{\,\prime}\, \rangle=
\langle \vec{p}\,|i\rangle \langle\vec{q}\,|\vec{q}^{\,\prime}\, \rangle=
(2\pi)^3\,\epsilon_i(\vec q\,)\,
\delta(\vec{q}-\vec{q}^{\,\prime}\,)\,f_i(\vec{p}\,)\quad\mbox{with}\quad
\epsilon_i(\vec q\,)= \left\{
\begin{array}{lcl}
2E_i(\vec q\,) & \mbox{for} & i=3 \\
\displaystyle\frac{E_i(\vec q\,)}{M_N} & \mbox{for} & i=1,2
\end{array}\right\}\,,
\end{equation}
where $f_i(\vec{p}\,)=\langle \vec{p}\,|i\rangle$ is the usual vertex function
of the separable representation.
For $i=3$ the Pauli principle for the nucleons is already incorporated, i.e.,
$P_{12}|3,\vec{q}\,\rangle=-|3,\vec{q}\,\rangle$ where $P_{12}$ is the nucleon
exchange operator.

The asymptotic channel wave function, describing the free motion of a particle
``$i$'' with momentum $\vec{q}$ relative to the interacting pair ($jk$),
is given by
\begin{equation}
|\phi_i(W,\vec{q}\,)\rangle = G_{\eta NN}(W)|i, \vec{q}\,\rangle\,,
\label{channelstate}
\end{equation}
where $G_{\eta NN}(W)$ is the free $\eta NN$ Green's function
depending on the total three-body energy $W$.
For the moment being we drop spin-isospin indices.
Then, expressing the separable $\eta N$- and $NN$-scattering matrices, acting
in three-particle space, in terms of the two--body matrix elements, we
find
\begin{equation}
\langle \vec{p}^{\,\prime}, \vec{q}^{\,\prime} |t_i(W)|
\vec{p}, \vec{q}\, \rangle =
(2\pi)^3 \delta(\vec{q}^{\,\prime}-\vec{q}\,)
\langle \vec{p}^{\,\prime}|t_i(W_i(W,\vec{q}\,))|\vec{p} \,\rangle =
(2\pi)^3 \delta(\vec{q}^{\,\prime}-\vec{q}\,) f^*_i(\vec{p}^{\,\prime})
\tau_i(W_i(W,\vec{q}\,))f_i(\vec{p}\,) \,,
\end{equation}
where the propagator of a pair ($jk$) in the presence of a spectator
``$i$'' reads
\begin{equation}
\tau_i(W_i)=\Big[\frac{1}{\gamma_i}-\frac{1}{(2\pi)^3}
\int\frac{d^3p}{\epsilon_j(\vec p\,)\,\epsilon_k(\vec p\,)}\,
\frac{|f_i(\vec p\,)|^2}{W_i-E_j(\vec p\,)-E_k(\vec p\,)+i\epsilon}
\Big]^{-1}\,.
\end{equation}
Here $W_i(W,\vec{q}\,)$ denotes the invariant mass of the subsystem
$(j,k)$, defined by putting the spectator particle
``$i$'' on mass shell, i.e.
\begin{equation}\label{Wi}
W_i(W,\vec{q}\,)=\sqrt{W^2-2\,W\,E_i(\vec{q}\,)+M_i^2}\,.
\end{equation}
For the particle energies we use the relativistic expressions
$E_i(\vec p\,)=\sqrt{p^2+M_i^2}$.
The separable ansatz leads to a system of
coupled equations of the familiar Lippman-Schwinger form \cite{Lovelace}
\begin{equation}
X_{ij}(W,\vec{q}^{\,\prime},\vec{q}\,)=Z_{ij}(W,\vec{q}^{\,\prime},\vec{q}\,)+
\sum\limits_{k=1}^3 \int \frac{d^3q^{\,\prime\prime}}{(2\pi)^3
\epsilon_k(q^{\,\prime\prime})}\,
Z_{ik}(W,\vec{q}^{\,\prime},\vec{q}^{\,\prime\prime})\,
\tau_k(W_k(W,\vec{q}^{\,\prime\prime}\,))\,
X_{kj}(W,\vec{q}^{\,\prime\prime},\vec{q}\,)\,.
\end{equation}
The amplitudes $X_{ij}(W)$ define the transitions between the
channel states (\ref{channel}), i.e.\ collisions of the
type ``$j + (ik) \to i + (jk)$'',
where $(ik)$ and $(jk)$ refer to interacting two-particle states.
The driving terms $Z_{ij}(W)$ are represented
by the matrix elements of the free $\eta NN$ Green's function
\begin{equation}
Z_{ij}(W,\vec{q}^{\,\prime},\vec{q}\,)=(1-\delta_{ij})
\langle i,\vec{q}^{\,\prime}|G_{\eta NN}(W)|j,\vec{q}\,\rangle\,.\label{Zij}
\end{equation}
Explicitly one finds for $i\ne j$
\begin{equation}\label{Zijex}
Z_{ij}(W,\vec{q}^{\,\prime},\vec{q}\,)=
\frac{f^*_i(\vec{p}_i(\vec{q}^{\,\prime},\vec{q}\,))\,
f_j(\vec{p}_j(\vec{q}^{\,\prime},\vec{q}\,))}{W-E_i(\vec q^{\,\prime})
-E_j(\vec q\,)-E_k(\vec{q}^{\,\prime}+\vec{q}\,)+i\,\epsilon}\,,
\end{equation}
where the momenta $\vec p_i(\vec{q}^{\,\prime},\vec{q}\,)$ and
$\vec p_j(\vec{q}^{\,\prime},\vec{q}\,)$ are given in terms of
$\vec q^{\,\prime}$ and $\vec q$. For simplicity we use the nonrelativistic
relations
\begin{equation}
\vec{p}_i(\vec{q}^{\,\prime},\vec{q}\,)=
\vec{q} +\frac{\mu_i}{M_k}\,\vec{q}^{\,\prime}\quad
\mbox{ and }\quad\vec{p}_j(\vec{q}^{\,\prime},\vec{q}\,)=
\vec{q}^{\,\prime}+\frac{\mu_j}{M_k}\,\vec{q}\,,
\end{equation}
where the reduced mass in $i$-th channel reads
\begin{equation}
\mu_i=\frac{M_jM_k}{M_j+M_k}\,.
\end{equation}

The next step to be taken towards an explicit evaluation of the three-body
equations is the
antisymmetrization of the basic amplitudes with respect to the exchange of the
nucleons  $N_1$ and $N_2$ for which we follow mainly the
work of~\cite{Afnan}. It affects only the channels $i=1$ and
$i=2$ because the channel $i=3$ is already antisymmetric
by construction as pointed out above. Consider the system of equations,
which couple the amplitudes $X_{ij}$ for the possible transitions from the
channel $j=3$. In the operator form we have explicitly
\begin{eqnarray}
X_{13}&=& Z_{13}+Z_{12}\,\tau_2\,X_{23}+Z_{13}\,\tau_3\,X_{33}\,,\nonumber\\
X_{23}&=& Z_{23}+Z_{21}\,\tau_1\,X_{13}+Z_{23}\,\tau_3\,X_{33}\,,
\label{X3equation}\\
X_{33}&=& Z_{31}\,\tau_1\,X_{13}+Z_{32}\,\tau_2\,X_{23}\,.\nonumber
\end{eqnarray}
Taking into account the identity of the nucleons, it is easy to find the
following relations~\cite{Afnan}
\begin{eqnarray}
\tau_1=\tau_2\,,\quad Z_{13}&=&-Z_{23}\,,\quad Z_{31}=-Z_{32}\,,\quad
\mbox{ and }\quad  Z_{12}=Z_{21}\,.
\label{Zsymmetry}
\end{eqnarray}
With the help of this symmetry one can reduce (\ref{X3equation}) to a
system of only two coupled equations
\begin{eqnarray}
(X_{13}-X_{23})&=&2Z_{13}-Z_{12}\,\tau_2\,(X_{13}-X_{23})
+2Z_{13}\,\tau_3\,X_{33}\,,\\
X_{33}&=&Z_{31}\,\tau_1\,(X_{13}-X_{23})\,.
\end{eqnarray}

Before defining the explicitly antisymmetrized amplitudes, it is convenient
to introduce a new channel notation. From now on we denote the channel
with a spectator nucleon as ``$N^*$'' and the one with a spectator meson as
``$d$''. The corresponding channel wave functions $|N^*,\vec{q}\,\rangle$ and
$|d,\vec{q}\,\rangle$ are assumed to be antisymmetrized with
respect to the nucleons, in detail
\begin{eqnarray}
|N^*,\vec{q}\,\rangle = \frac{1}{\sqrt{2}}\,
(|1,\vec{q}\,\rangle - |2,\vec{q}\,\rangle)\quad \mbox{ and }\quad
|d,\vec{q}\,\rangle = |3,\vec{q}\,\rangle\,.
\end{eqnarray}
Defining the driving terms in a symbolic notation by
\begin{equation}\label{DefZij}
Z_{N^*N^*}=-\frac{1}{2}(Z_{12}+Z_{21})=-Z_{12}\,,\quad
Z_{dN^*}=Z_{31}\,, \quad \mbox{ and }\quad Z_{N^*d}=Z_{13}\,,
\end{equation}
and the properly antisymmetrized amplitudes by
\begin{equation}
X_{d}=X_{33}\,, \quad
X_{N^*d}=\frac{1}{2}\,(X_{13}-X_{23})\,,
\end{equation}
one arrives at the following set of equations
\begin{eqnarray}
X_{N^*d}&=&Z_{N^*d}+Z_{N^*d}\,\tau_d\,X_{d}+Z_{N^*N^*}\,\tau_{N^*}\,X_{N^*d}\,,\\
X_{d}&=&2\,Z_{dN^*}\,\tau_{N^*}\,X_{N^*d}\,,
\end{eqnarray}
where the amplitudes $X_{d}$ and $X_{N^*d}$ describe the two different
transitions $\eta d\to \eta d$ and  $\eta d\to N^*N$, respectively,
which are realized in $\eta d$ scattering.

Now we will consider in addition the
coupling to the $\pi NN$ channel via the two-body reaction $\eta N\to
\pi N$, whereas we will neglect the coupling to the two-pion channel
$\pi\pi NN$. Its inclusion into the three-body formalism would
require the use of phenomenological approaches, which in any case seem
to be very ambiguous. Due to the smallness of the $N^*\to\pi\pi N$ decay
probability we believe that this neglect will not significantly influence
our results.
A treatment of the resulting coupled channel problem within the
Faddeev approach was developed, e.g., in~\cite{Ued84,Miyagawa}.
Accordingly, we extend the channel $|N^*\rangle$ to the following
two-component form
\begin{equation} |N^*\rangle=\left( \begin{array}{cc}
|N^*(\eta)\rangle \\
|N^*(\pi)\rangle
\end{array}
\right)\,.
\end{equation}
The corresponding coupled channel $t$-matrix is given by
\begin{equation}\label{tN}
t_{N^*}=|N^*\rangle \,\tau_{N^*} \,\langle N^*|\,,
\end{equation}
with the $N^*$-propagator
\begin{equation}\label{Npropgtr}
\tau_{N^*}(W_{N^*})=\Big[\frac{1}{\gamma_{N^*}}-\frac{1}{(2\pi)^3}\,
\sum\limits_{\alpha\in\{\pi,\eta\}}
\int\frac{M_N}{2\,E_{\alpha}(\vec p\,)\,E_N(\vec p\,)}\,
\frac{|f^{(\alpha)}_{N^*}(\vec p\,)|^2}
{W_{N^*}-E_N(\vec p\,)-E_{\alpha}(\vec p\,)+i\epsilon}\,d^3p\,\Big]^{-1}\,,
\end{equation}
where $f^{(\alpha)}_{N^*}(\vec p\,)=\langle \vec p\,|N^*(\alpha)\rangle$.
Turning now to the three-body problem, we obtain a set of three coupled
equations, namely
\begin{eqnarray}\label{ThrBEq}
X_{N^*d}&=&Z^{(\eta)}_{N^*d}+
Z^{(\eta)}_{N^*d}\,\tau^{(\eta)}_d\,X^{(\eta)}_d+
Z^{(\pi)}_{N^*d}\,\tau^{(\pi)}_d\,X^{(\pi)}_d+
(Z^{(\eta)}_{N^*N^*}+Z^{(\pi)}_{N^*N^*})\,
\tau_{N^*}\,X_{N^*d}\,,
\label{XNeq}\\
X_{d}^{(\eta)}&=&2\,Z^{(\eta)}_{dN^*}\,\tau_{N^*}\,X_{N^*d}\,,\label{Xdeq1}\\
X_{d}^{(\pi)}&=&2\,Z^{(\pi)}_{dN^*}\,\tau_{N^*}\,X_{N^*d}\,,\label{Xdeq2}
\end{eqnarray}
where the driving terms are given in analogy to (\ref{DefZij}) for
$\alpha\in\{\pi,\eta\}$ by
\begin{eqnarray}
Z^{(\alpha)}_{N^*N^*}=
-\frac{1}{2}(Z^{(\alpha)}_{12}+Z^{(\alpha)}_{21})\,,\quad
Z^{(\alpha)}_{dN^*}=Z^{(\alpha)}_{31} \quad \mbox{ and }\quad
Z^{(\alpha)}_{N^*d}=Z^{(\alpha)}_{13}\,,
\end{eqnarray}
with analogous definitions for $Z^{(\alpha)}_{ij}$ as in (\ref{Zij}), i.e.
\begin{equation}
Z^{(\alpha)}_{ij}(W,\vec{q}^{\,\prime},\vec{q}\,)=(1-\delta_{ij})
\langle i,\vec{q}^{\,\prime}|G_{\alpha NN}(W)|j,\vec{q}\,\rangle\,.
\label{Zijalpha}
\end{equation}
The set of equations (\ref{XNeq}) through (\ref{Xdeq2}) is the formal basis
of the present calculation. Its
solution gives the required symmetrized rearrangement amplitudes and thus
amounts to solving the $\eta NN$-problem.

Now we will specify the separable $\eta N$- and $NN$-scattering matrices
which determine the driving two-body forces.
Since we work entirely in the low-energy region we shall neglect all but the
$S_{11}$-$\eta N$-interaction. Analogously, only $s$-wave $NN$-states
($^1S_0,\,^3S_1$) are included in view of their strong dominance. As a
first step, we omit the tensor part of the nucleon-nucleon force, because
their inclusion would introduce further calculational complications.
Since we restrict the pairwise interactions to $s$-waves only,
the vertex functions have a very simple structure
\begin{equation}\label{eqFF}
\langle \vec p\,|k\rangle=f_k(\vec p\,)=g_k\,F_k(\vec p\,)
\,,\quad\mbox{ with }\quad
F_k(\vec p\,)=\frac{\beta_k^2}{\beta_k^2+(\vec p\,)^2}\,, \quad
k\in \{d,N^*\}\,.
\end{equation}
For the $s$-wave $NN$-scattering matrix
\begin{equation}\label{eqTNN}
\langle \vec{p}^{\,\prime}|t_d(W_d)|\vec{p} \,\rangle =
f^*_d(\vec p^{\,\prime}) \,\tau_d(W_d)\, f_d(\vec p\,)\,,
\end{equation}
the following parametrization has been used
\begin{equation}\label{eqGNN}
g_d^2=\frac{16\pi a}{a\beta_d-2}\,, \quad
\gamma_d=-\frac{1}{2M_N}\,,\label{vertexpar}
\end{equation}
where $a$ denotes the corresponding $NN$-scattering length.
In order to simplify the numerical evaluations we reduce the function
$\tau_d$ to the nonrelativistic form
\begin{equation}\label{eqTaud}
\tau_d(E_{NN})=-\frac{1}{2M_N}\,
\Big[1+\frac{g_d^2\beta_d^3}
{16\pi\Big(i\beta_d+\sqrt{M_NE_{NN}}\Big)^2}
\Big]^{-1}\,,
\end{equation}
with $E_{NN}$ being the kinetic $NN$-energy in their center of mass.
To be more consistent one can, e.g., apply the minimal relativity to the
separable nonrelativistic $NN$-amplitude, but we have checked that the
results are not affected by the approximation (\ref{eqTaud}).

The $NN$-interaction parameters were taken from a fit of
low-energy $np$-scattering~\cite{Yamag}, yielding
\begin{equation}
\beta_d=1.4488\,\mbox{fm}^{-1},
\quad a=\left\{\begin{array}{rl}
5.378\,\mbox{fm}& \mbox{for the $^3S_1$-state}\cr
-23.690\,\mbox{fm}& \mbox{for the $^1S_0$-state}\cr
\end{array}\right\}\,.
\end{equation}

For the $N^*$ $t$-matrix (\ref{tN}) we use the energy dependent coupling
strength $\gamma_{N^*}=(W_{N^*}-M_0)^{-1}$ so that the $N^*$-propagator is
given in the familiar isobar form
\begin{equation}\label{eqTauN}
\tau_{N^*}(W_{N^*})=(W_{N^*}-M_0-\Sigma_\pi(W_{N^*})-
\Sigma_\eta(W_{N^*})+i\,\epsilon)^{-1}\,,\label{N*prop}
\end{equation}
The self energy contributions $\Sigma_\alpha(W_{N^*})$
$(\alpha\in\{\pi,\eta\})$ from the $\pi N$- and $\eta N$-loops are defined
by the equation (\ref{Npropgtr}).  The meson-$N^*$ vertices are
parametrized by
\begin{equation}\label{eqFi}
f^{(\alpha)}_{N^*}(\vec p\,)
=g^{(\alpha)}_{N^*}\, F_{N^*}^{(\alpha)}(\vec p\,)\,, \quad
\mbox{with}\quad F_{N^*}^{(\alpha)}(\vec p\,)
=\frac{\beta_{N^*}^{(\alpha)2}} {\beta_{N^*}^{(\alpha)2}+(\vec
p\,)^2}\,.
\end{equation}
The small partial decay into the $\pi\pi N$-channel has been neglected for
consistency.

In the actual calculation we have chosen the following set of
$N^*$-parameters
\begin{equation}\label{eqBeTa}
g_{N^*}^{(\eta)}=2.0\,,\ g_{N^*}^{(\pi)}=1.1\,,\
\beta_{N^*}^{(\eta)}=6.5\,\mbox{fm}^{-1}\,, \
\beta_{N^*}^{(\pi)}=4.5\,\mbox{fm}^{-1}\,, M_0=1622\, \mbox{MeV}\,,
\end{equation}
which gives for the $\eta N$-scattering length
\begin{equation}\label{aEtaN}
a_{\eta N}=(0.75+0.27i)\,\mbox{fm}^{-1}\,.
\end{equation}
This value was obtained in \cite{Wycech} using a
phenomenological analysis of the coupled channels $\pi N,\, \pi\pi N,\,
\eta N$ and $\gamma N$, and is in close agreement with other
recent results for low-energy $\eta N$-scattering \cite{Bat1}.

In order to complete the formal part, we only have to generalize our
formalism to
add also spin-isospin degrees of freedom. Since we retain only
$s$-wave orbitals for the two-body interactions, excluding in particular
tensor forces, the total spin $S$ is a good quantum number. Thus,
we obtain the partial wave decomposition of the basic states as
\begin{equation}
|SM_S, TM_T, \vec{q}\, \rangle =
|(\sigma_i\sigma_j)S_k\sigma_k; SM_S \rangle
|(\tau_i\tau_j)T_k\tau_k; TM_T \rangle
\,4\pi\sum\limits_{LM_L} |q, LM_L \rangle Y_{LM_L}(\hat{q})
\label{basis}
\end{equation}
in a self explanatory notation for the coupling of the individual spins and
isospins to $S$ and $T$, respectively. In the basis of (\ref{basis}) the Born
amplitudes are given by $(i,j\in\{d,N^*\})$
\begin{equation}
Z_{ij}(W,\vec{q}^{\,\prime},\vec{q}\,)=4\pi\,\Lambda^{ST}\,
\sum\limits_L (2L+1)Z^L_{ij}(W,q^{\prime},q)P_L(\hat q'\cdot \hat q)\,,
\end{equation}
where the driving term $Z_{ij}^L$ of the partial wave $L$ is
defined by
\begin{equation}\label{ZLij} Z^L_{ij}(W,q',q)= \frac{1}{8\pi}
\int\limits_{-1}^{1}
Z_{ij}(W,\vec{q}^{\,\prime},\vec{q}\,)\,P_L(\hat q'\cdot \hat q) \,
d(\hat q'\cdot \hat q)\,.
\end{equation}
The recoupling coefficients
\begin{equation}
\Lambda^{ST}=\langle (\sigma_i\sigma_j)S_k\sigma_k, S|
(\sigma_j\sigma_k)S_i\sigma_i, S \rangle \
\langle (\tau_i\tau_j)T_k\tau_k, T|
(\tau_j\tau_k)T_i\tau_i, T \rangle
\end{equation}
are evaluated using the standard formula (see e.g.~\cite{Edmonds})
\begin{equation}
\langle (\sigma_i\sigma_j)S_k\sigma_k, S|
(\sigma_j\sigma_k)S_i\sigma_i, S \rangle=
(-1)^{\sigma_j+\sigma_k-S_i}\sqrt{(2S_i+1)(2S_k+1)}
\left\{
\begin{array}{ccc}
\sigma_i & \sigma_j & S_k \\
\sigma_k & S        & S_i
\end{array}
\right\}\,,
\end{equation}
and the analogous expression for the isospin recoupling.
After a partial wave decomposition we have a
system of coupled integral equations in only one variable, which
schematically reads
\begin{equation}
X^{LST}=\Lambda^{ST}\,Z^L+\Lambda^{ST}\,Z^L\,\tau\, X^{LST}\,.
\end{equation}
In the case of $\eta d$ scattering, only states with
$T=0$ and $S=1$ contribute. But in $\eta$-photoproduction on the deuteron,
other $ST$-combinations are realized.

In the Appendix we describe briefly the techniques involved to invert the
system (\ref{XNeq})-(\ref{Xdeq2}). The corresponding mathematical apparatus
is quite standard by now,
and we are concerned with some formal aspects only, connected with the
relativistic kinematics explored in this paper.

\section{\mbox{\boldmath$\eta \lowercase{d}$} elastic scattering}\label{scattering}

\begin{figure}
\centerline{\psfig{figure=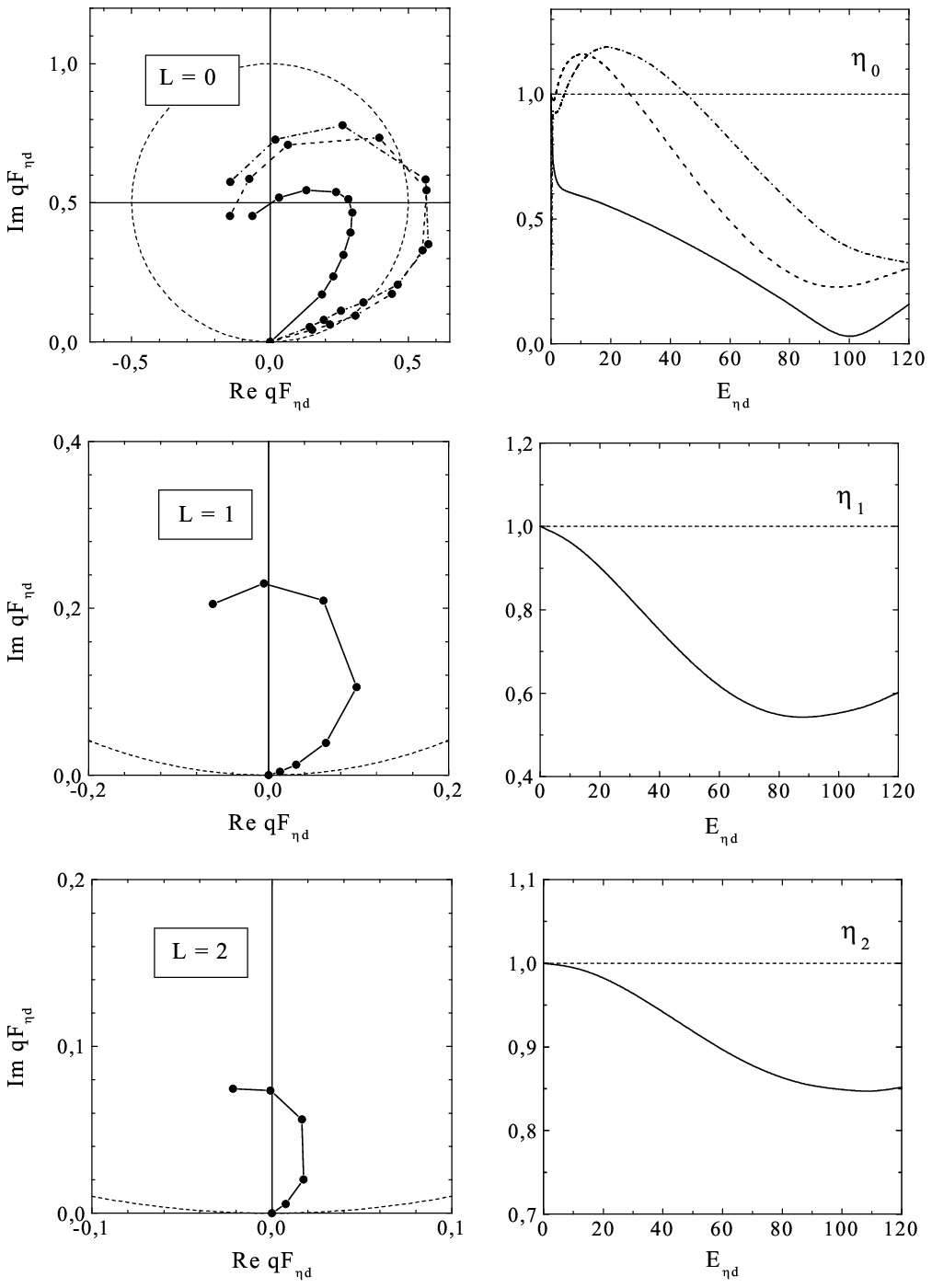,width=12cm,angle=0}}
\caption{
Argand diagrams (left panels) and inelasticity parameters (right panels)
for elastic $\eta d$-scattering for the lowest $\eta d$ partial waves
$L=0,1$, and 2 as a function of the c.m.\ kinetic energy $E_{\eta d}$.
The dots mark the following energies for $L=0$: $E_{\eta d}=0$, 0.5, 1, 2,
4, 8, 16, 32, 64, 90, 120 MeV; for $L=1$: $E_{\eta d}=0$, 4, 8, 16, 32, 64,
90, 120 MeV; and for $L=2$: $E_{\eta d}=0$, 16, 32, 64, 90, 120
MeV; For $L=0$, the dashed curve shows the result from the first order
rescattering approximation and the dash-dotted one the result of the
optical model. For the other partial waves, the first order rescattering
approximation coincides essentially with the complete calculation.  }
\label{fig1}
\end{figure}

\begin{figure}
\centerline{\psfig{figure=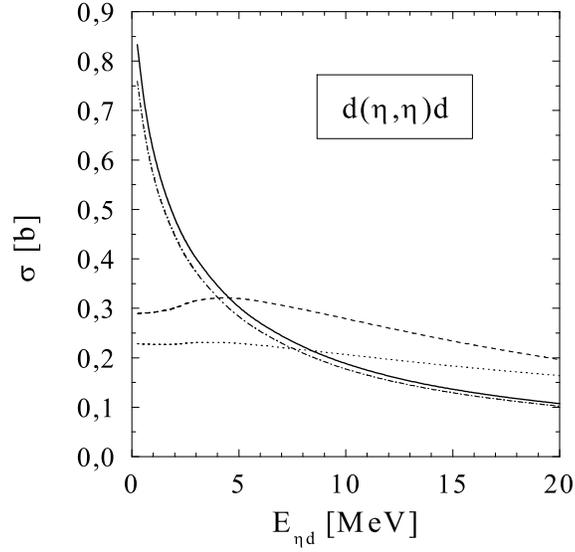,width=8cm,angle=0}}
\caption{
Elastic $\eta d$-total cross section versus the c.m.\ kinetic energy
$E_{\eta d}$ for various theoretical ingredients.  Notation of the curves:
dotted: IA; dashed: first order rescattering; solid: complete three-body
calculation; dash-dotted: three-body calculation but
without $\pi$-exchange contribution to the driving term $Z_{N^*N^*}$.
}
\label{fig2}
\end{figure}

The scattering amplitude is determined by the matrix element
$X_d^{(\eta)}$ of (\ref{XNeq})-(\ref{Xdeq2}) in the ($S=1,T=0$)-channel
\begin{equation}
F_{\eta d}^L(q)=-\frac{E_d}{4\pi W}\,N_d^2\,X_d^{(\eta)L}(W,q,q)\,,
\end{equation}
with $q$ being the on-shell $\eta d$ c.m.\ momentum.
The factor $N_d$ takes into account the normalization of the deuteron wave
function to unity.
In our parametrization (\ref{vertexpar}) we have
\begin{equation}
N_d^2=8\,\pi\,\frac{\epsilon_d^2}{g_d^2}
\Big(\frac{1}{\beta}+\frac{1}{\sqrt{M_N\,|\epsilon_d|}}\Big)^3\,,
\end{equation}
where $\epsilon_d$ is the deuteron binding energy.
For the c.m.\ differential cross section one has the usual expression
\begin{equation}
d\sigma(\eta d\rightarrow\eta d)=\left|4\pi
\sum\limits_L(2L+1)\,F_{\eta d}^L(q)\,P_L(\cos\theta)\,\right|^2 d\Omega\,.
\end{equation}
We have calculated the scattering amplitudes for the first
three partial waves, $L=0$, 1, and 2. The results are summarized in
Fig.~\ref{fig1}, where the Argand plots as well as the corresponding
inelasticity parameters $\eta_L$ are presented.
The following conclusions may be drawn:

(i) The rather rapid increase of the s-wave amplitude close to
the scattering threshold is explained
by the presence of a virtual pole in the ($S=1,T=0$) $\eta d$-state.
This pole has been located in \cite{FiAr00} on the nonphysical two-body sheet
of the $\eta d$ c.m.\ kinetic energy plane. As is noted in the introduction, 
the existence of the virtual state leads to a strong enhancement of the
scattering cross section, which is presented in Fig.~\ref{fig2}.

(ii) With increasing energy the Argand plots show a resonance-like behaviour
around the position of the elementary $N^*$-resonance. These $\eta d$
pseudoresonances are explained simply by the spreading of the elementary
$\eta N$-interaction
over different partial waves when viewed from the $\eta d$ c.m.-system.
In this energy region the amplitudes become highly inelastic, so that
the scattering is almost diffractive. In the $\eta$-production reactions this
effect appears as a strong absorption of the produced mesons inside the
nucleus, leading to the so-called surface production mechanism \cite{R.Landau}.

(iii) The $N^*N$-interaction generated by pion exchange is almost
negligible. This smallness is well understood if one takes
into account the dominance of the low momentum part of the
$N^*N$-interaction at low energy in on-shell $\eta d$ scattering.
We expect that due to the smallness of the pion mass the
exchange of the retarded pion determines mainly the high momentum
component of $Z_{N^*N^*} $ in (\ref{XNeq})
\begin{equation}\label{Mesonaustsch}
Z_{N^*N^*} \equiv Z^{(\eta)}_{N^*N^*}+Z^{(\pi)}_{N^*N^*}
\end{equation}
and thus is not very important in our case.  In this connection, we find
quite puzzling the recent results of \cite{Pena2} where the authors claim a
very strong sensitivity of the $\eta d$ scattering to the contribution of
$\pi$-exchange, which visibly weakens the strength of the $\eta d$
attraction and results in a drastic reduction of the elastic cross section.
The effect of pion exchange in $\eta d$ elastic scattering as
predicted by our model is shown in Fig.~\ref{fig2}.  In contrast
to~\cite{Pena2}, it is positive and quite small.
One essential distinction from our work is that very different cut-off
parameters $\beta_{N^*}^{(\eta)}\approx 13\,\mbox{fm}^{-1}$ and
$\beta_{N^*}^{(\pi)}=1.2\,\mbox{fm}^{-1}$ are used in \cite{Pena2}. One could
suspect that in this case the $\pi N$-force may become important due to
its relatively long-range nature. In order to check the sensitivity of the
results to the choice of the meson-nucleon cut-offs we have
performed a calculation with a new set of $N^*$-parameters including the 
same cut-off parameters as in \cite{Pena2}
\begin{equation}\label{eqBeTa1}
g_{N^*}^{(\eta)}=2.13\,,\ g_{N^*}^{(\pi)}=3.8\,,\
\beta_{N^*}^{(\eta)}=13\,\mbox{fm}^{-1}\,, \
\beta_{N^*}^{(\pi)}=1.2\,\mbox{fm}^{-1}\,, M_0=1656\, \mbox{MeV}\,.
\end{equation}
In this case, without pion exchange one observes a reduction
of the total cross section by about 30$\%$ close to zero
energy and by about 10$\%$ at $E_{\eta d}=20$\,MeV, which remains
much different to the results of~\cite{Pena2}. 

At this point one may speculate about another possible explanation: 
A further distinction from our work is, that in~\cite{Pena2}  
nonrelativistic kinematics for all participating
particles was used. In this case, in going from $\eta$- to $\pi$-exchange
one finds in the expression of the denominator
of the driving term $Z_{N^*N^*}^{(\pi)}(W,\vec{p}^{\,\prime},\vec{p})$
(see (\ref{Zij}), (\ref{Zijex}) and (\ref{Zijalpha}))
an additional mass difference $M_\eta-M_\pi$, i.e.
\begin{equation}
W-E_N(\vec p_i)-E_N(\vec p_j)-\sqrt{(\vec{p}_i+\vec{p}_j)^2+M_\pi^2}
\to E+M_\eta-M_\pi-\frac{p_i^2}{2M_N}-\frac{p_j^2}{2M_N}-
\frac{(\vec{p}_i+\vec{p}_j)^2}{2M_\pi}\,,
\end{equation}
with $E=W-2M_N-M_\eta$. Examining the formalism in~\cite{Pena} and 
\cite{Pena2}, one possibility could be 
that this term was absent in the calculation. It is clear that the
mass difference shifts effectively the on-shell kinetic energy to higher
values, $E\to E+M_\eta-M_\pi \gg E$, so that
only the less important high momentum part of the $N^*N$-interaction becomes
sensitive to $\pi$-exchange.

(iv) In order to show to which extent the first order rescattering terms
alone take into account the rescattering effect, we have also plotted their
contributions to the Argand plots and have presented the corresponding cross
section in Fig.~\ref{fig2}. We see that the major contributions
beyond the first order terms are the ones to the $L=0$ partial wave amplitude.
This is
consistent with our intuitive expectation that the higher order terms in a
multiple scattering series involve essentially the short distances in the
interacting
system, which contribute mostly to the $s$-wave part.
This fact is demonstrated by the Argand plots, where we see that the $s$-wave
amplitude is very poorly represented by the first order term. In the region
$E_{\eta d}=2-20$\,MeV the letter goes beyond the unitary circle.
On the other hand, the
first order approximation is well justified for higher partial waves with
$L=1$ and 2.
This effect is also partially explained by the specific features
of our two-body input where only the $s$-wave orbitals are involved.

It would be useful to check the correct three-body results against those
given by the optical model. This
model, in the KMT version, was applied to the description of the final state
interaction in the $\gamma d\to \eta d$ reaction \cite{Kamal97}.
Its crucial approximation is the neglect of
target excitations in between scatterings. The domains within which this
adiabatic idea is expected to work is, for example, pion scattering on
medium and heavy nuclei. But it is not clear whether the optical picture is
applicable to the $\eta NN$-system, because of its nonadiabatic nature this
approach may be too restrictive. Therefore, we consider here this
approximation in order to establish the range of its validity and to
understand its connection with such a fundamental model as the Faddeev
theory. In order to arrive at the appropriate equations from our
three-body ansatz (\ref{XNeq})-(\ref{Xdeq2}), we neglect the terms,
containing $Z^{(\alpha)}_{N^*N^*}$ and change also
the function $\tau_d$ of (\ref{eqTaud}) to the pure pole form
\begin{equation}
\tau_d=\frac{N_d^2}{E_{NN}+|\epsilon_d|}\,.
\end{equation}
For the transition amplitude $X_{d}$, we then obtain the equation
\begin{equation}
X_{d}=V_d+\frac{1}{2}V_d\,\tau_d\,X_{d}\,,\label{optmod}
\end{equation}
where
\begin{equation}
V_d=2Z_{dN^*}\,\tau_{N^*}\,Z_{N^*d}
=\langle\phi_d|\sum\limits_{i=1,2}t_{\eta N}(i)|\phi_d\rangle
\end{equation}
is the familiar first order optical potential for $\eta d$ scattering.
Following the authors of~\cite{Kamal97} we have introduced the additional 
factor 1/2 in (\ref{optmod}) in order to avoid the double counting of the
$\eta N$-interaction. The $s$-wave amplitude $F^0_{\eta d}(q)$
obtained within this approach is also shown in Fig.~\ref{fig1}.
We see that
the quality of the optical model prediction is rather poor notably near the
scattering threshold. It underestimates distinctly the $\eta NN$-interaction
strength at low energies. Furthermore neglecting the excitation of the
$NN$-subsystem breaks the three-body unitarity so that the corresponding
inelasticity parameter exceeds unity.
We may conclude that due to the limitations discussed above, the
optical model is unable to incorporate the important properties of
the $\eta NN$ interaction near threshold and is a rather bad
approximation to the exact theory.

\section{Break up channels and unitarity}\label{unitarity}

Up to now, the break up reactions $\eta d\to \eta np$ and $\eta d\to \pi NN$
have not been investigated in detail, one of the reasons being the much more
complex
calculations involved. The transition matrix element $X_0$ for the break up
process $\eta d\to \eta np$ may easily be evaluated once
the rearrangement amplitudes $X^{(\eta)}_{d}$ and $X_{N^*d}$ in the
($S=1,T=0$) channel are calculated for
the appropriate region of the final phase space. Then one has
\begin{eqnarray}\label{eqX0}
X_0(W,\vec{p}_1,\vec{p}_2,\vec{q},\vec{k}\,)
&=&(f_{N^*}^{(\eta)}(\vec{p}_{\eta 1})\,
\tau_{N^*}(W_{N^*})\,X_{N^*d}(W,\vec{p}_2,\vec{k})+ (1\leftrightarrow 2))
+ f_d(\vec{p}_{1 2})\,
\tau_d(W_d)\,X^{(\eta)}_{d}(W,\vec{q},\vec{k}\,)\,,\label{breakup}
\end{eqnarray}
where the amplitudes $X_{N^*d}$ and $X^{(\eta)}_{d}$ can be decomposed
according to
\begin{equation}
X(W,\vec{p},\vec{p}^{\,\prime}\,)=4\pi
\sum\limits_L(2L+1)\,X^L(W,p,p')\,P_L(\hat{p}\cdot\hat{p}')\,.
\end{equation}
Here $W$ denotes the total three-body c.m.\ energy and $\vec{k}$
the momentum of the incident $\eta$-meson. The three-momenta of the
final nucleons and the $\eta$-meson are denoted by $\vec{p}_1$, $\vec{p}_2$
and $\vec{q}$, respectively. The arguments of the vertex functions
$f_i(\vec{p}\,)$ in (\ref{breakup}) are the nonrelativistic two-body relative
momenta, determined by the final state kinematics as ($i=1,2$)
\begin{equation}
\vec{p}_{\eta i}=\frac{M_N\,\vec{q}-m_\eta\,\vec{p}_i}{M_N+m_\eta}\,,\ \quad
\vec{p}_{12}=\frac{1}{2}\,(\vec{p}_1-\vec{p}_2)\,.
\end{equation}
The c.m.\ break up cross section is then given by
\begin{equation}\label{crsec}
d\sigma(\eta d\rightarrow \eta np)
=\frac{1}{(2\pi)^5}\,\frac{E_d(\vec{k}\,)M_N^2}{4W\,k}\,
N_d^2\,
|X_0(W,\vec{p}_1,\vec{p}_2,\vec{q},\vec{k}\,)|^2
\,d\Omega_\eta \,dE_\eta \,d\phi_{N_1}
\,dE_{N_1}\,,
\end{equation}
where the initial deuteron energy is $E_d(\vec{k}\,)=\sqrt{k^2+M_d^2}$.
The angle $\phi_{N_1}$ is the azimuthal angle of
the nucleon momentum $\vec{p}_1$ relative to the $\vec k$-$\vec q$ plane in
the frame where the $z$-axis is chosen along $\vec{q}$.  For the absorptive
channel $\eta d\to \pi NN$, only the first two terms in (\ref{breakup})
remain as follows from the spin-isospin selection rules.

\begin{figure}
\centerline{\psfig{figure=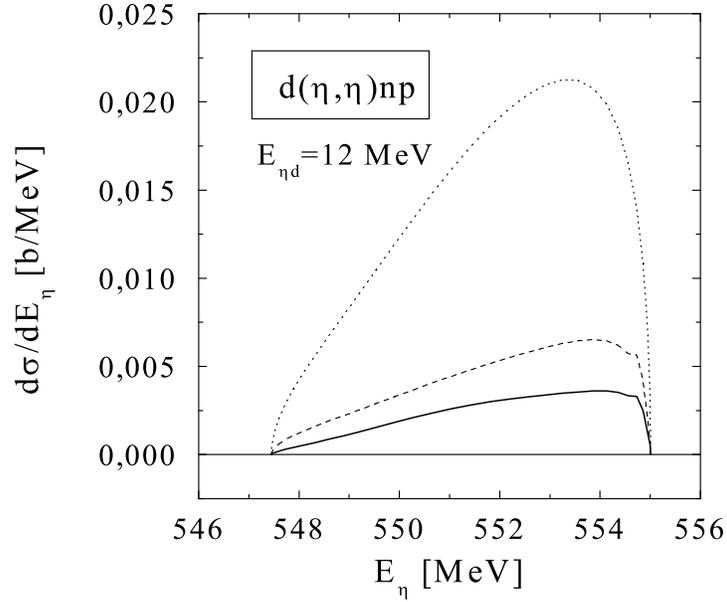,width=10cm,angle=0}}
\caption{
Spectrum of the emitted $\eta$-mesons in inelastic $\eta d$ scattering
calculated in the $\eta d$ c.m.\ system
for an initial c.m.\ kinetic energy $E_{\eta d}$=12 MeV.
Notation of the curves: dotted : IA; dashed: first
order rescattering; solid: complete three-body calculation.  }
\label{fig3}
\end{figure}

Our predictions for the $\eta$-meson spectrum are shown in Fig.~\ref{fig3}.
Firstly, one observes a strong decreases of the cross section
by the $\eta NN$ interaction of about 85 $\%$. This effect may be explained,
at least partially, by the orthogonality of the initial and final $NN$ wave
functions. Also within the first order rescattering approximation one finds
a strong reduction of the IA, but the size of this effect is
evidently underestimated, so that this approximation results in a cross
section about a factor of 2 larger than the complete calculation.

\begin{figure}
\centerline{\psfig{figure=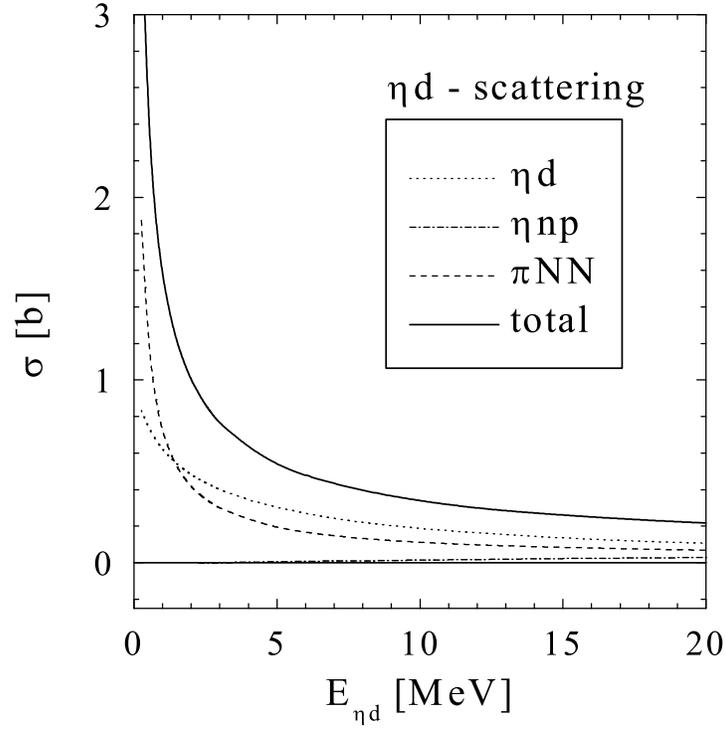,width=10cm,angle=0}}
\caption{
Various contributions to the total $\eta d$-cross section
from the different channels. The solid curve shows the sum of all channels.
}
\label{fig4}
\end{figure}

The total cross sections for the elastic and inelastic channels are shown in
Fig.~\ref{fig4}.  We see that the $\eta d\to \eta np$ cross section is in
general small, also due to the damping effect noted above. On the other
hand, the absorptive channel $\eta d\to \pi NN$ gives an essential part of
the total cross section. This may be explained firstly by the
relatively large phase space available for pion emission already at low
initial kinetic energies. Secondly, the cross section diverges as $1/v$,
where $v$ is the $\eta d$ relative velocity, near the elastic scattering
threshold, which is typical for exothermic reactions.

\begin{figure}
\centerline{\psfig{figure=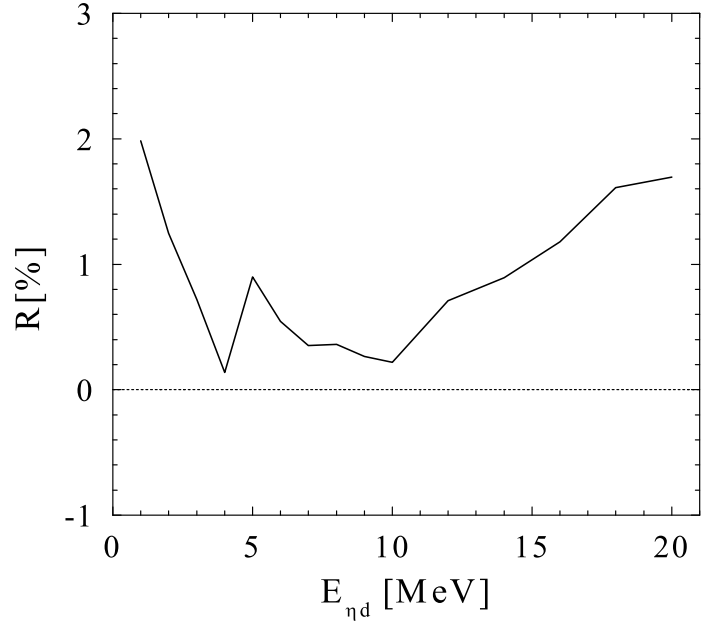,width=10cm,angle=0}}
\caption{
Relative deviation $R$ between the total cross sections calculated either
explicitly from summing the various reaction channels or from the optical
theorem (see definition in (\protect{\ref{dev}})).
}
\label{fig5}
\end{figure}

The break up processes provide us also with a check of the question whether
the limitation to the lowest
partial waves is justified by considering the unitarity relation, which
couples these channels to the elastic one. An important
consequence from this relation is the well known optical theorem,
which in our case reads
\begin{equation}
\frac{4\pi}{q}\Im m F_{\eta d}(\theta=0)=\sigma_{tot}=\sigma(\eta d\to\eta
d) +\sigma(\eta d\to\eta np)+\sigma(\eta d\to\pi NN)\,.\label{opticalth}
\end{equation}
Thus a comparison of both sides of this relation allows one to estimate
how crucial the truncation
of the partial wave expansion is.
Furthermore, this is also a natural method to check the accuracy of the
numerical procedure applied for solving the basic equations
(\ref{XNeq})-(\ref{Xdeq2}).
To this end we have calculated both
sides of (\ref{opticalth}) and present in Fig.~\ref{fig5} the relative
deviation
\begin{equation}\label{dev}
R=\frac{\frac{4\pi}{q}\Im
m\,F_{\eta d}(\theta=0)-\sigma_{tot}}{\sigma_{tot}}\,.
\end{equation}
The fact, that $|R|$ does not exceed 2$\%$ over the whole energy range
considered, gives us
confidence that the approximations of the present approach are not crucial
and that our results are well founded. But one has to keep in mind that at
higher energies the neglected higher partial waves have to be included and
very likely also the two-pion channel which has been neglected completely
in this work.

\section{\mbox{\boldmath$\eta$}-photoproduction on the deuteron}\label{photoproduction}

We now will turn to $\eta$-production reactions which are more easily
accessible in an experiment. Our main concern will be to see to what extent the
strong final-state $\eta NN$-interaction will influence the dynamical
properties of such reactions and what type of information on the
$N^*N$-interaction can be extracted from them. Most interesting are
the coherent and incoherent $\eta$-photoproduction processes
$\gamma d\to \eta d$ and $\gamma d\to\eta np$ which, for the reasons noted in
the introduction, were extensively investigated both theoretically and
experimentally during recent years.

As usual, we treat the electromagnetic interaction in first order
perturbation theory. To apply the formalism represented
by the system in (\ref{XNeq})-(\ref{Xdeq2}) to the $\eta$-photoproduction,
one just has to replace the $\eta$-meson by the photon in the entrance
channel.  This then yields a set of coupled equations which are formally
similar to (\ref{XNeq})-(\ref{Xdeq2}) but where the driving interaction in
(\ref{XNeq}) has been changed according to
\begin{equation}
Z^{(\eta)}_{N^*d}\to \langle N^*(\gamma)|\,G_{\gamma NN}\,|d \rangle\,,
\end{equation}
where $G_{\gamma NN}$ denotes the free propagator in the
$\gamma NN$-sector and $|N^*(\gamma)\rangle$ the
electromagnetic vertex function for the transition $\gamma N\to N^*$.
The method of inversion of the corresponding three-particle equations,
described in the Appendix, remains of course the same. For the electromagnetic
vertex $|N^*(\gamma)\rangle$ we use the parametrization
\begin{equation}
\langle \vec{p}\,|N^*(\gamma)\rangle\
=g_{\gamma NN^*}^{(N)}\frac{\beta_{N^*}^{(\gamma)2}}
{\beta_{N^*}^{(\gamma)2}+k_{\gamma N}^2}\,,
\end{equation}
where $k_{\gamma N}$ is the $\gamma N$ c.m.\ momenta and
$\beta_{N^*}^{(\gamma)}$ was taken to be 480 MeV.
In the proton channel
the value $g_{\gamma NN^*}^{(p)}$ was determined by fitting the
experimental $\gamma p\to \eta p$ cross section \cite{Krusche1}. For the
neutron channel we have used the relation
$g_{\gamma NN^*}^{(n)}=-0.80\,g_{\gamma NN^*}^{(p)}$ which is compatible
with different analyses \cite{Krusche2,Hoff,Hejny}. In principle,
there is a complex phase between the proton and neutron amplitudes as has
been pointed out in~\cite{RiAr00}. However, the incoherent reaction is not
sensitive to this phase and thus its neglect is not crucial there. On the
other hand, this phase is important for the strength of the isoscalar
part of the amplitude which determines completely the coherent reaction.
This is discussed below in Sect.~\ref{coherent}.

Keeping in mind the discussion given in Sect.~\ref{scattering},
we can expect that the higher order
terms of the multiple scattering series are important only for the lowest
partial waves of the transition amplitude and have very little effect on the
higher partial waves. This
allows us to treat the latter perturbatively by taking into
account the first order rescattering terms only. This
may easily be done by adding and subtracting the first order
rescattering amplitude from that given by the full three-body calculation.
Schematically we may write
\begin{equation}\label{X_XR}
X=X^{resc}+(X-X^{resc})\,.
\end{equation}
The first term can be calculated directly using the
cartesian basis without making a partial wave decomposition.
The corresponding techniques have been described before, e.g.,
in~\cite{FiAr97,Hoshi,Halder}, and need not be repeated here.
The
second term in (\ref{X_XR}) converges very rapidly to zero for the higher
partial waves, so that only the contribution from the $L=0$ part has to be
taken into account.

In the $\gamma d$ c.m.\ system, the cross sections for the coherent and
incoherent photoproduction reactions are
\begin{eqnarray}
d\sigma(\gamma d\to \eta d)
&=&\frac{q}{k}\frac{E_d(\vec{k}\,)E_d(\vec{q}\,)}
{(4\pi W)^2}\,\frac{N_d^2}{6}\sum\limits_{\mbox{\tiny spins}}
|X^{(\eta)}_d(W,\vec{q},\vec{k}\,)|^2\,d\Omega_\eta\,,\label{cohxsection}\\
d\sigma(\gamma d\to \eta np)
&=&\frac{1}{(2\pi)^5}\,\frac{E_d(\vec{k}\,)M_N^2}{4W\,k}\,
\frac{N_d^2}{6}\sum\limits_{\mbox{\tiny spins}}
|X_0(W,\vec{p}_1,\vec{p}_2,\vec{q},\vec{k}\,)|^2
\,d\Omega_\eta \,dE_\eta \,d\phi_{N_1}
\,dE_{N_1}\,,
\label{incohxsection}
\end{eqnarray}
where all notations are the same as in (\ref{crsec}) with $\vec{k}$ being
the photon momentum.

From the results obtained for $\eta d$-scattering
we can expect that the strong $\eta NN$-interaction will appreciably influence
the observables of $\eta$-photoproduction, too. Indeed, the strong
attraction in the $\eta NN$-system, already mentioned above, tends to hold the
participating particles in the region where the primary ``photoproduction
interaction'' works. Since the rate of the ($\gamma \to \eta$)-transition will
be proportional to the probability of finding the
produced $\eta$ in this region,
a strong enhancement of the $\eta$-production cross section may be
expected near threshold~\cite{Watson}.

\subsection{Coherent \mbox{\boldmath$\eta$}-photoproduction on the deuteron}\label{coherent}

\begin{figure}
\centerline{\psfig{figure=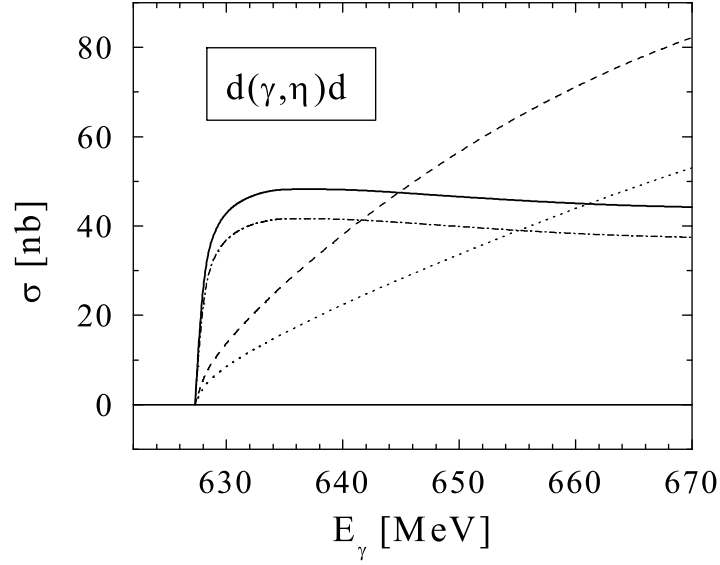,width=10cm,angle=0}}
\caption{
Total cross section for the coherent reaction $\gamma d\to\eta d$.
Notation of the curves:
dotted: IA;
dashed: first order rescattering;
solid:  complete three-body calculation;
dash-dotted: three-body calculation
without $\pi$-exchange contribution.
}
\label{fig6}
\end{figure}

The coherent channel $\gamma d\to \eta d$ has been investigated
theoretically rather extensively in contrast to very few reliable
experimental data reflecting the enormous difficulty for separating this
small cross section from the dominant incoherent reaction.
One of the most important questions in relation to the coherent
photoproduction is the isotopical separation of the $\gamma N\to N^*$
amplitude as already pointed out above. Recent work~\cite{RiAr00}
has shown that the consideration of the relative phase between the proton
and neutron $N^*$-photoexcitation amplitudes is very essential.
Consequently, it is practically impossible to extract the isoscalar part
$g^{(s)}_{\gamma NN^*}$ of the $\gamma N\to N^*$ amplitude
from the observation of only  $\gamma p\to \eta p$ and $\gamma d\to \eta np$
reactions. The model dependent fixing of this phase
may be done, e.g., by the analysis of the multipole pion photoproduction
amplitudes, as in \cite{RiAr00,BeTa}. Since this question is beyond the
scope of the present paper, we treat the modulus of the isoscalar part of
the amplitude as a free parameter.
Our predictions for the total $\gamma d\to \eta d$ cross section
obtained with
\begin{equation} \alpha=\left|\frac{g^{(s)}_{\gamma
NN^*}}{g^{(p)}_{\gamma NN^*}}\right|=0.26
\end{equation}
are presented in Fig.~\ref{fig6}. One readily notes that the strong
attraction in the $(S=1,T=0)$-channel reflects itself in a drastic increase
of the cross section over a mere IA-calculation just above threshold. It is
worth mentioning that the experimental confirmation of this result, i.e.,
the observation of the strong enhancement of the $\eta$-yield close to the
threshold would imply that the present theoretical ideas on the $\eta NN$
low-energy dynamics are substantially correct.  In view of the strong
s-wave dominance in the $\gamma d\to \eta d$ cross section near threshold,
we may expect an important influence of the higher order rescattering
terms. As is demonstrated in Fig.~\ref{fig6}, taking into account only
first order rescattering, we are not able to reproduce neither the common
trend nor the size of the results given by the complete calculation.
Compared to the recent work of~\cite{RiAr00}, the present results confirm
the inadequacy of the first order rescattering.
However, the energy dependence of the total cross section
differs substantially from those of~\cite{RiAr00}.
This difference is of course a consequence of the present three-body
approach to the $\eta NN$-system which was beyond the scope
of~\cite{RiAr00}.

Another feature which deserves a comment is that, due to the
large momentum transfer associated with the large $\eta$-mass,
the high momentum
part of the $N^*N$-interaction becomes important already near threshold. The
latter leads to the increase of the $\pi$-exchange
contribution which was found to
be almost negligible in $\eta d$-scattering
(see the discussion in Sect.~\ref{scattering}). As we can see, the general
effect of including the $\pi$-exchange is to enhance the near-threshold
cross section by more than a factor 1.2.
This sensitivity of the $\gamma d\to \eta d$ reaction to the pion
contribution was also noted in \cite{RiAr00}. We must furthermore
mention that the
importance of the $N^*N$ rearrangement potential $Z_{N^*N^*}$
(\ref{Mesonaustsch}) in the $\pi$- and
$\eta$-channels depends strongly on the
strength of the $\eta NN^*$ and $\pi NN^*$ couplings used.

In comparison with the results of~\cite{Shev2}, we firstly would like to
emphasize once more that in our case the source of the strong final state
interaction is the virtual $\eta NN$ s-wave pole found in \cite{FiAr00}.
Therefore it is not surprising that our theory leads to substantially
different results than those of~\cite{Shev2}
where the s-wave $\eta d$-resonance governs the whole reaction dynamics. In
particular, we found no evidence from the resonant peak in
the total cross section at $E_\gamma=635$ MeV.
It should also be noted that the effect of $\eta d$-interaction obtained by
us, although
being essential, is however not as strong as predicted in the pioneering
work \cite{Ued92}.

\begin{figure}
\centerline{\psfig{figure=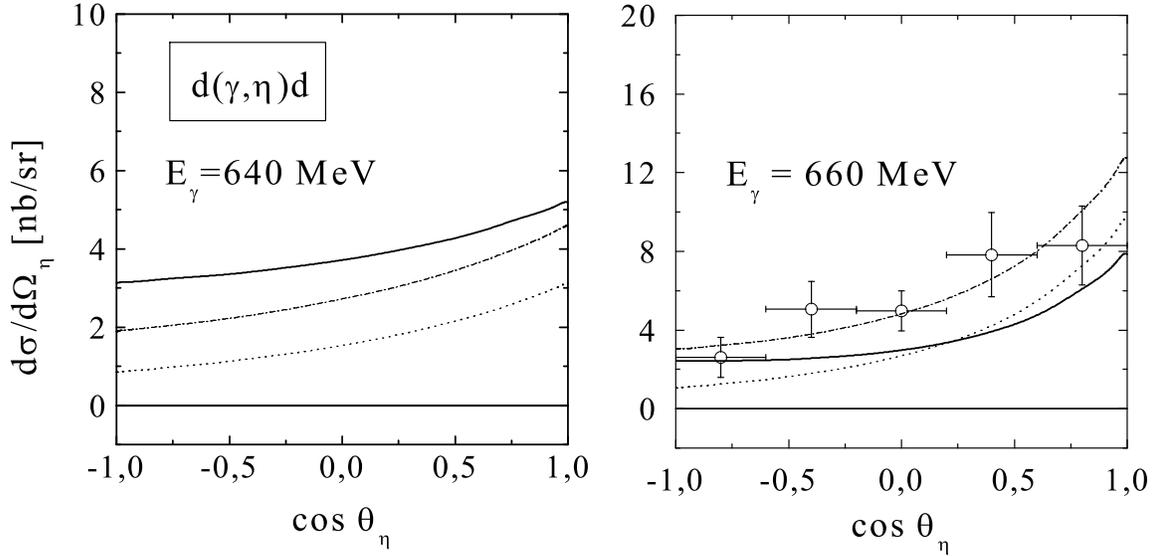,width=16cm,angle=0}}
\caption{
Differential cross section for the coherent reaction $\gamma d\to\eta d$
in the $\gamma d$ c.m.\ system
at two lab photon energies $E_\gamma = 640$ and 660~MeV. Notation of the
curves:  dotted: IA; dashed: first order rescattering; solid: complete
three-body model.
The open circles represent the data for an energy bin $E_\gamma =
652-664$~MeV from~\protect\cite{Hoff}.  }
\label{fig7}
\end{figure}

In Fig.~\ref{fig7} we show our results for the differential cross section at
two photon energies.
Analyzing the influence of the $\eta
NN$-interaction we conclude that the final state interaction tends to
enhance the $s$-wave part of the reaction amplitude and thus to make the
angular distribution somewhat more isotropic.
The magnitude of the available experimental data \cite{Hoff}
covering the near-threshold region is not
reproduced with $\alpha=0.26$ but one must note that the results of
the most recent measurements \cite{Metag,Weiss} are consistent with this
value of $\alpha$.

\subsection{Incoherent channel \mbox{\boldmath$\gamma d\to \eta np$}}

With respect to the incoherent $\eta$-photoproduction on the
deuteron, there exist fewer theoretical studies as for the coherent
reaction. In our previous work on $\gamma d\to \eta np$~\cite{FiArPL,FiAr97},
we have addressed the effects of final state rescattering. Besides the
investigation of the main properties of the cross section in the near
threshold region, we have studied in particular the importance of three-body
dynamics in the final $\eta np$ system. It has been shown that a restriction
to the first order rescattering with respect to the $NN$- and $\eta N$-final
state interaction does not give a sufficiently accurate approximation to the
$s$-wave reaction amplitude and that higher order terms make very substantial
contributions. At this point we would like to note that the numerical results
in~\cite{FiArPL} contained an error. The correct calculation presented here
yields somewhat larger predictions. In this work, we also would like
to complement our previous studies.

\begin{figure}
\centerline{\psfig{figure=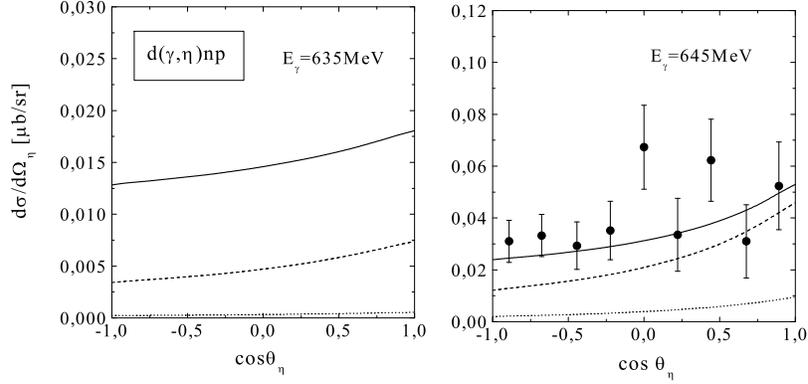,width=11cm,angle=0}}
\caption{
Angular distribution of $\eta$-mesons of the incoherent
reaction $\gamma d\to\eta np$
calculated in the $\gamma d$ c.m.\ system
at the two lab photon energies.
Notation of the curves as in \protect\ref{fig7}.
The experimental points are the inclusive
$\gamma d\to\eta X$ measurements from \protect\cite{Krusche2}.
}
\label{fig8}
\end{figure}

As first, we show in Fig.~\ref{fig8} the angular
distribution of the outgoing meson, calculated in the c.m.\ system at the
energy 5 and 15 MeV above threshold. One readly sees that
the strong attraction in the s-wave $(S=0,T=1)$ state
implies a substantial enhancement over the IA cross section while
making the angular distribution rather isotropic.
As was already discussed in our previous paper~\cite{FiArPL}, the first
order rescattering is unable to describe close to threshold the size of the
cross section obtained within the three-body formalism.

\begin{figure}
\centerline{\psfig{figure=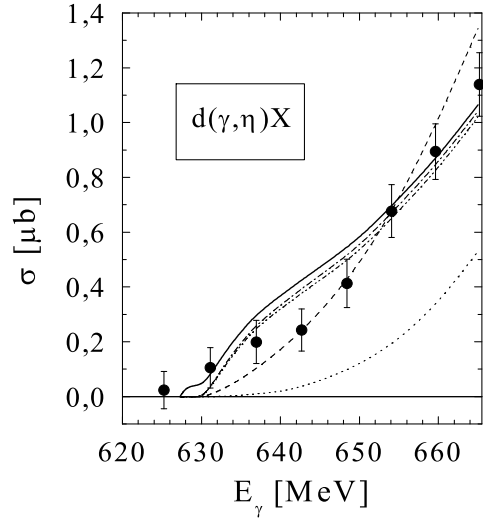,width=7cm,angle=0}}
\caption{
Total cross section for the reaction $\gamma d\to\eta X$. Notation
of the curves: dotted: IA; dashed: first order rescattering;
dash-double-dot: complete the three-body model;
dash-dot: three-body model
without the contribution of $\pi$-exchange; solid: sum of coherent and
incoherent channels.
Inclusive $\gamma d\to\eta X$ data are taken from \protect\cite{Krusche2}.
}
\label{fig9}
\end{figure}

The importance of the $\eta NN$-interaction for the
$\eta$-photoproduction reaction is most apparent in the total cross
section exhibited in Fig.~\ref{fig9}. It was shown in~\cite{FiAr97} that
the impulse approximation (IA)
is strongly suppressed because of the large spectator nucleon momenta
required for the production of a low-energy $\eta$-meson in this
approximation. In contrast to the IA, the complete three-body approach
results in a cross section which exhibits a strong enhancement starting
right from threshold. Therefore, the observed rather high values of the
$\eta$-meson yield near threshold are explained in principle.
Adding the contribution from the coherent channel, we obtain the inclusive
cross section $\sigma(\gamma d\to \eta X)=\sigma(\gamma d\to \eta np)+
\sigma(\gamma d\to \eta d)$ which was actually measured in the
experiment~\cite{Krusche2}. One readily notices, that in this region
our theoretical predictions are in reasonable agreement with the data.

In the same figure we demonstrate the role of pion exchange in the break-up
channel. The dash-dotted curve shows the results obtained when the
terms containing $Z_{N^*d}^{(\pi)}$, $Z_{dN^*}^{(\pi)}$ and 
$Z_{N^*N^*}^{(\pi)}$
in equations (\ref{XNeq}) through (\ref{Xdeq2}) are neglected.
In this case, in contrast to the coherent reaction, the role of the
neglected terms appears to be rather small. This is due mainly to the fact
that the pion exchange potential $Z_{N^*N^*}^{(\pi)}$ in the $T=1$ channel
is three times weaker then in the $T=0$ state. The role of the other two terms
$Z_{N^*d}^{(\pi)}$ and $Z_{dN^*}^{(\pi)}$, which do not contribute to the
coherent process, turns out to be totally negligible.

Here we would like to note that our primary goal was
to treat as precisely as possible the three-body aspects of the problem,
which requires, of course, a few simplifications of some model ingredients.
In particular, we have neglected here the tensor
$NN$-force and consequently the D-wave component of the deuteron wave
function. As was shown in~\cite{FiAr97}, due to the
large momentum transfer, the
contribution arising from the D-wave part of the deuteron shows up
near threshold and reduces the pure S-wave cross section sizeably.
Therefore, a more refined analysis should be based on a more realistic
deuteron wave function. 

\section{Conclusions and outlook}\label{conclusion}

In the present paper, which we consider as an exploratory step towards a
quantitative understanding of the dynamical properties of the
$\eta NN$-interaction at low energy, we have shown that in view of the
relatively strong attraction, both in the $\eta N$- and
$NN$-subsystems, the appropriate theoretical framework is given by
the three-body scattering approach of Faddeev type. It allows one to
include all orders of the rescattering expansion systematically.
We were interested mainly in the energy region
of low kinetic energies with particular emphasis on the first few MeV above
the $\eta d$-scattering threshold, which is of special interest for
$\eta$-nuclear studies.
Our main results may be summarized as follows:

(i) In the low-energy regime,
the major contributions of higher order rescattering
beyond the first order term appear in the $L=0$ partial wave.
This observation is consistent with the notion
that higher order terms involve mainly short distances in the deuteron.
Our explicit study shows that for the higher partial waves there is no need
to invert the three-body equations, since the corresponding Neumann series
converges very rapidly, so that only the first nontrivial term is sufficient
in order to take into account rescattering effects in these partial waves.

(ii) The importance of rescattering in the inelastic channel
$\eta d\to \eta np$ is a necessary consequence of
the orthogonality of the initial and final
$NN$-states. We have shown that this effect largely reduces the probability of
the deuteron break up in the allowed kinematical domain.

(iii) The $\eta
NN$ dynamics, as manifest
in the photon induced reactions $\gamma d\to \eta d$ and
$\gamma d\to \eta np$, is to some extent unrelated to
the on-shell $\eta d$-interaction. In particular, the exchange of retarded
pions, which is found to be negligible in elastic scattering, becomes
much more pronounced in the coherent photoproduction process
because of the increased importance of the high momentum components
of the $N^*N$-interaction.

(iv) The exact treatment of the three-body aspects of the $\eta NN$ final
state in the $\eta$-photoproduction on a deuteron
is quite essential for a quantitative understanding
of the form as well as the size
of the angular distributions of produced etas.
The general effect of the $\eta NN$-interaction is to enhance
the $s$-wave part of the scattering amplitude close to the production
threshold. As a consequence, we found a distinct shift of the major
part of the $\eta$-meson yield to the
low-energy region as well as a rather isotropic character of the corresponding
angular distributions.

Finally, we would like to make a few remarks with respect to the use of the
developed theory for $\eta$-photoproduction on heavier nuclei. Firstly, we
would like to recall that close to threshold this reaction is accompanied by a
large momentum transfer $\Delta p$ in the whole region of available
emission angles. On the other hand, the energy $\Delta E$ deposited into the
nucleus is minimal, such that
\begin{equation}
\Delta E \ll \frac{(\Delta p)^2}{2M_N} \,.
\end{equation}
Therefore, one may expect that the single particle response
preferring the ``quasifree'' kinematics, where
\begin{equation}
\Delta E \approx \frac{(\Delta p)^2}{2M_N}\,,
\end{equation}
must be strongly
suppressed. We have already observed this effect in the reaction
$\gamma d\to\eta np$, where a strong suppression of the IA-contribution
was found. Thus in this situation the mechanism where two ore more
nucleons are allowed to share the transferred momentum becomes important.
As the dominant reaction mode we thus may assume a two-nucleon response where
the $\eta$-meson is produced on a correlated pair of nucleons. The case
of three or more participating particles requires the presence of more than
two nucleons close together which seems to be less probable. Furthermore,
the strong attraction of the $\eta NN$-system as found in the present work
tends to hold it together, so that the
production of a strongly correlated $NN^*$-pair and its decay into the $\eta
NN$-channel may appear to be the most important mechanism for
$\eta$-photoproduction on nuclei near threshold. Taking into account these
qualitative arguments, we may conclude that our approach may also be useful for
the study of $\eta$-photoproduction on complex nuclei.

\acknowledgments
A.F.\ is grateful to the theory group for the kind hospitality during his 
stay at the Institute of Nuclear Physics of the University of Mainz. 

\renewcommand{\theequation}{A\arabic{equation}}
\setcounter{equation}{0}
\section*{Appendix: Method of solution}
\label{app}

In order to solve the equations (\ref{XNeq})-(\ref{Xdeq2}),
we use the method of contour deformation developed mainly in~\cite{Heth}
for elastic scattering and extended for break-up processes in~\cite{AAY}.
We will not give any details at this point and consider only that part of
the formalism, which is connected with relativistic kinematics which is not
touched upon in the literature to the best of our knowledge. The
method was developed to avoid the problems,
caused by the moving logarithmic singularities in the terms
$Z_{ij}^L(W,q',q)$. As is well known, these singularities
arise from the exchange of a particle with positive kinetic energy. Above the
inelastic threshold (in our case we have two inelastic thresholds -- $\pi
NN$ at $W\approx 2017$~MeV and $\eta NN$ at $W\approx 2425$~MeV), 
$Z_{ij}^L$ contains a right-hand cut and, as a consequence, becomes complex,
where the imaginary part is as usually determined by the discontinuity
across the cut. As an example, let us consider a driving term $Z_{ij}^L$ 
of the general form
\begin{equation}
Z_{ij}^L(W,q',q)=\frac{1}{8\pi}\int\limits_{-1}^{1}
\frac{f^*_i(\vec p_1(\vec{q}^{\,\prime},\vec{q}\,))\,
f_j(\vec p_2(\vec{q}^{\,\prime},\vec{q}\,))}
{W-E_i(\vec{q}^{\,\prime})-E_j(\vec{q}\,)
-\sqrt{(\vec{q}^{\,\prime}+\vec{q}\,)^2+M^2_k}+i\epsilon}
P_L(\hat q'\cdot \hat q)\, d(\hat q'\cdot \hat q)\,,\quad
i,j,k\in\{d,N^*\}\,.
\label{app_pot}
\end{equation}

\begin{figure}
\centerline{\psfig{figure=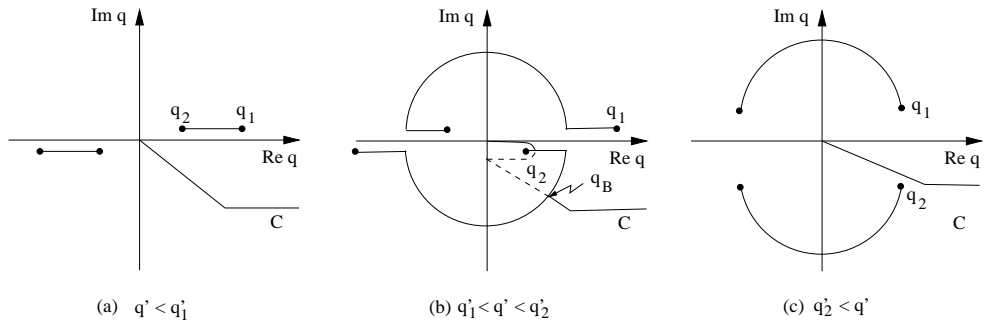,width=13cm,angle=0}}
\vspace{.5cm}
\caption{
Deformed integration path $C$ and location of the cuts of the
rearrangement potential $Z_{ij}(W,q',q)$ in the complex $q$-plane
for three different regions of the momentum $q'$.
Note that for case (b) the integration path crosses two times the cut
so that in between $C$ lies in the nonphysical sheet as indicated by the
dashed line.
}
\label{fig10}
\end{figure}

The main types of singularities of $Z_{ij}^L$
as a function of $q$ for three different regions of the momentum $q'$
are shown in Fig.~\ref{fig10}.
For the boundary values of $q^{\prime}$ one finds
\begin{eqnarray}
q^{\prime}_1&=&
\frac{\lambda^{1/2}\left((W-M_j)^2,M_i^2,M_k^2\right)}{2(W-M_j)}
\label{Q1bndr}\\
q^{\prime}_2&=&\frac{\lambda^{1/2}\left(W^2,M_i^2,(M_j+M_k)^2\right)}{2W}
\label{Q2bndr}\,,
\end{eqnarray}
where the triangle function $\lambda(x,y,z)$ has the form
(see, e.g., \cite{Byck})
\begin{equation}
\lambda(x,y,z)=\left(x-(\sqrt{y}+\sqrt{z})^2\right)
\left(x-(\sqrt{y}-\sqrt{z})^2\right)\,.
\end{equation}

The position of the logarithmic poles $q_1$ and $q_2$ can easily be found
to be
\begin{eqnarray} q_1&=&\frac{\alpha q'+D}{W_i^2}\,,\\
q_2&=&\left\{
\begin{array}{ll}
\displaystyle\frac{-\alpha q'+D}{W_i^2}\,, &\mbox{for}\quad q'< q'_1\\
\displaystyle\frac{\alpha q'-D}{W_i^2}\,, &\mbox{for}\quad q'> q'_1
\end{array}
\right\},
\end{eqnarray}
where $W_i$ is defined in (\ref{Wi}) and
\begin{eqnarray}
\alpha&=&\frac12(M_j^2-M_k^2+W_i^2)\,,\\
D&=&(W-E_i(\vec{q}\,'))\sqrt{\alpha^2-M_j^2W_i^2}\,.
\end{eqnarray}

Also shown in Fig.~\ref{fig10} is the deformed contour $C$ along which
$Z_{ij}^L(W,q',q)$ as a function of $q$ for fixed $q'$
has no singularities so that the equations (\ref{XNeq})-(\ref{Xdeq2})
may be solved by matrix inversion.
The solutions for real values of $q'$ may be obtained by
the one-fold iteration of the same equations
along the real axis $q$. In the case
of pure elastic processes, such as $\eta d\to \eta d$ and $\gamma d\to \eta d$,
only the case (a) may be realized and the iteration becomes trivial. Some
problems arise for the break up processes  $\eta d\to \eta np$ and $\gamma
d\to \eta np$, where also the case (b) is realized in the appropriate region
of $q'$. In this case, the knowledge of the amplitude $X$ in the interval
$[0,q_2]$ is
needed. By direct manipulation, one can see that the inequality
$q^{\prime}_{1\,min}>q_2$ always holds. Here
$q^{\prime}_{1\,min}$ is the minimal value
from the boundary momenta $q^{\prime}_1$
obtained according to (\ref{Q1bndr})-(\ref{Q2bndr}) for all potentials
$Z_{ij}$ involved in our dynamical equations (\ref{XNeq}) --
(\ref{Xdeq2}). Therefore, the needed values of the $X$-matrix are
always known and the case (b) leads only to a non-essential calculational
complexity. In this case the iteration is carried out according to the
formula (omitting the spin-isospin notations)
\begin{equation}\label{CaseB}
Z_{ij}^L\,\tau_j\,X^L_j=\frac{2}{\pi}\int\limits_0^{q_2}
Disc\,[Z^L_{ij}(W,q',q'')]\,\tau_j(W_j)\,X^L_j(W,q'',q')
\frac{q''^2dq''}{\epsilon_j(q'')}
+\frac{2}{\pi}
\int\limits_C Z^L_{ij}(W,q',q'')\tau_j(W_j)X^L_j(W,q'',q')
\frac{q''^2dq''}{\epsilon_j(q'')}\,,
\end{equation}
where, as one can prove using the definitions of (\ref{Zijex}) and 
(\ref{ZLij})
\begin{equation}
Disc\,[Z^L_{ij}(W,q',q'')]=i\pi\frac{g_ig_j\beta_i^2\beta_j^2}{16\pi}
\frac{M_k^2}{\mu_i\mu_j}\frac{P_L(x_0)}
{(q'q'')^3(a_i+x_0)(a_j+x_0)}\,,
\end{equation}
with
\begin{eqnarray}
x_0&=&\frac{(W-E_i(q')-E_j(q''))^2-M_k^2-q'^2-q''^2}{2q'q''}\,,\\
a_i&=&\frac{M_k}{2q'q''\mu_i}\,(\beta_i^2+q''^2+(q'\mu_i/M_k)^2)\,,\\
a_j&=&\frac{M_k}{2q'q''\mu_j}\,(\beta_j^2+q'^2+(q''\mu_j/M_k)^2)\,.
\end{eqnarray}
It is important that when integrating along the contour $C$ in the
case (b) one has to go into the second sheet (dashed line between
the points 0 and $q_B$ in Fig.~\ref{fig10}).
In the case (c) $q^{\prime} > q^{\prime}_2$, the singularities are shifted
into the complex plane and therefore do not cause any problems.

In conclusion we would like to note, that the
kinematical area in which the case (b) is realized is rather small. As
the direct calculation shows, the neglect of the first term in
(\ref{CaseB}) does not lead to any markable change of the results.

\newpage

\vspace{-.5cm}

\end{document}